\documentclass[12pt]{article}

\advance\hoffset by -1.1truecm
\advance\voffset by -1.0truecm

\def\be{\begin{equation}}
\def\ee{\end{equation}}
\def\ba{\begin{eqnarray}}
\def\ea{\end{eqnarray}}

\newcommand{\R}{\mbox{I \hspace{-0.82em} R}}
\newcommand{\ve}{{\bf t}}
\newcommand{\sn}{\smallskip\newline}
\newcommand{\mn}{\medskip\newline}

\newcommand{\tphi}{{\tilde{\phi}}}
\newcommand{\B}{ {B_{\omega_{max}}} }
\newcommand{\Hi}{ {H_{\omega_{max}}} }

\begin{document}
\title{On Fields with Finite Information Density}
\author{Achim Kempf\\
Departments of Applied Mathematics and Physics, University of Waterloo\\
Perimeter Institute for Theoretical Physics\\
Waterloo, Ontario, Canada\\
{\small Email: akempf@uwaterloo.ca}}

\date{}

\maketitle
\begin{abstract}
The existence of a natural ultraviolet cutoff at the Planck scale is
widely expected. In a previous Letter, it has been proposed to model this
cutoff as an information density bound by utilizing suitably generalized
methods from the mathematical theory of communication. Here, we prove the
mathematical conjectures that were made in this Letter.
\end{abstract}
\section{Introduction} 
Gedanken experiments indicate that the basic spatial notions of length,
area and volume lose operational meaning at the Planck scale of
$10^{-35}m$ (assuming 3+1 dimensions). The argument is that any attempt at
resolving smaller scales would imply a momentum uncertainty large enough
to cause a curvature uncertainty which would spoil the attempted spatial
resolution. It has been widely suggested, therefore, that the uncertainty
relations should contain corrections that imply a finite lower bound to
the achievable uncertainty in localizing the position or time of events,
see \cite{qg}. If there exists such a smallest length, area or volume in
nature this should imply some form of an upper bound to the information
density that can be imprinted on physical fields. This is indicated as
well by studies of holography, see e.g. \cite{bekensteinetc}.

The question arises, however, in which way such an upper bound to
information density could be modelled mathematically. In \cite{ak-s}, it
was proposed to model the information density bound in essentially the
same way that has been successful in standard information theory and its
applications to communication engineering.

In the early days of communication engineering, the problem arose how to
quantify the amount of information contained in continuous signals of
music, speech or images. Claude Shannon solved the problem in two steps in
his seminal work \cite{shannon} which started modern information theory.
First, he noticed that bandlimited continuous signals such as music are
fully captured by discrete samples, i.e. that they are fully
reconstructible from those samples, if the spacing of those samples does
not exceed $1/2\omega_{max}$, where $\omega_{max}$ is the bandwidth of the
signal, i.e., the largest frequency that occurs in the signal. The
mathematical discipline which studies the reconstruction of functions from
samples is now known as Shannon sampling theory and this is what we here
attempt to generalize for applications to Planck scale physics. The
proposal is that physical fields could be captured everywhere if sampled
only at discrete points in space, as long as the spacing of those sampling
points is sufficiently small, say of the order of the Planck distance. As
the second step, Shannon considered the information density limiting
effect of noise in the signal. While important, we will here not consider
this second issue, other than mentioning that the role of noise could be
played by quantum fluctuations.

In order to generalize Shannon sampling theory for use in the physics of
general curved spacetimes, it is important to develop a sampling theory
that allows both the bandwidth and the density of samples to vary
continuously: the original Shannon sampling theorem requires
equidistantly-spaced samples but on generic curved spaces there are no
equidistantly spaced lattices.

In \cite{ak-s}, as a first step, a generalization of Shannon sampling
theory for the case of one dimension was outlined and its main features
were conjectured. On the basis of those conjectures, a new method of data
compression that utilizes variable sampling rates (adapted to the signal)
was developed and a patent has been granted, \cite{ak-patent}. Here, we
will explicitly carry out the program which was outlined in \cite{ak-s},
thereby proving the conjectures that were made in \cite{ak-s}. In
addition, we show that we are dealing with a reproducing kernel Hilbert
space and we explicitly calculate its reproducing kernel.

Recently, in \cite{ak-covsam}, this framework has been generalized to
describe a covariant informat\-ion-theoretic cutoff in
arbitrary-dimensional curved spacetime. The results and techniques
developed in the present paper should be very useful for explicitly
carrying out the much more general program proposed in \cite{ak-covsam}.

\section{Terminology and notation} 
We are concerned with the theory of continuous functions that can be
reconstructed from any discrete set of samples that is sufficiently
tightly spaced. This field of research, sampling theory, is
well-established both in mathematics and in communication engineering.
While we are here targeting applications to Planck scale physics we will
use the established terminology. The translation of terms is obvious:
signals mean real scalar fields, bandwidth means ultraviolet cutoff, the
time coordinate for signals could be both a time coordinate or a space
coordinate for fields, etc. Note that in the case of fields, unlike in the
case of signals, samples of field amplitudes are generally not directly
measurable. Both in quantum mechanics and in quantum field theory the
measurement problem is deep and unsolved. If discrete samples suffice to
capture a field everywhere this may have profound implications on the
measurement problem. While very interesting, we will here not further
pursue this question. Also, we will be concerned with the sampling theory
of scalar fields only. The generalization to vector, spinor and
operator-valued fields will be considered elsewhere.

Let us begin by noting that the mathematical language of the generalized
sampling theory that we are concerned with is also the mathematical
language of quantum mechanics, namely functional analysis. While this
suggests the use Dirac's bra-ket notation, there are good reasons for
using the proper mathematical notation instead. For example, instead of
Dirac's $\langle \phi \vert A\vert \psi\rangle$ we will write $(\phi,A
\psi)$. The reason is that we will have to be precise about the domains
and adjoints of operators. While an expression such as $(A\phi,\psi)$ is
unambiguous, the same expression using Dirac's notation would require
bracketing which becomes exceedingly cumbersome in more complicated
situations.

\section{Shannon sampling theory}
We begin by recalling elements of Shannon sampling theory. The basic
Shannon sampling theorem states that in order to capture a signal
$\phi(t)$ with bandwidth $\omega_{max}$ for all times $t$, it is
sufficient to record only the signals' values at the discrete points in
time $\{t_n\}$ with spacing $s= t_{n+1}-t_{n}$, where:
\be
s= \frac{1}{2\omega_{max}}
\ee
If the samples are taken at this rate, which is the so-called Nyquist rate,
then the  Shannon sampling formula allows the reconstruction of
the continuous signal $\phi(t)$ at \it all \rm times
$t$ from its values $\phi(t_n)$ at the discrete times $\{t_n\}$:
\begin{equation}
\phi(t)~ = ~\sum_{n=-\infty}^{+\infty}
~\frac{\sin\left(2\pi~(t-t_n)~\omega_{max}
 \right)}{2\pi~(t-t_n)~\omega_{max}} ~~\phi(t_n)
\label{e1}
\end{equation}
There is a 1-parameter family of ``sampling time lattices" $\{
t_n{(\alpha)} \}$ which all have the required spacing $s$, namely \be
t_n(\alpha )~  =~ \frac{n}{2\omega_{max}} +\alpha \label{sisp} \ee where
$\alpha \in [0,1/2\omega_{max}]$. The sampling formula Eq.\ref{e1} applies
equally to all of these sampling lattices. \mn The sampling theorem has a
simple proof which combines discrete and continuous Fourier transforms.
Indeed, the basic idea of  the sampling theorem is reported \cite{marks}
to go back as far as to Borel (1897), or even to Cauchy (1841). The
theorem was introduced to information theory by Shannon in  1949, see
\cite{shannon}. In the meanwhile, numerous generalizations of the theorem
have been worked out. For example, in practical communication engineering
applications one usually applies a generalization of the sampling theorem
which improves the convergence of the sampling expansion at the price of a
slight oversampling, typically on the order of $10 \%$. For standard
references see e.g. \cite{jerri,marks}. For recent work see, e.g.,
\cite{zayed,seip,annaby} and references therein. The ability to completely
recover continuous functions from discrete samples, through the sampling
theorem, is being applied ubiquitously in fields ranging from music on CDs
and scientific data taking to pure mathematics where it is used, for
example, to turn certain sums into easier-to-evaluate integrals.

Our aim here, as outlined in \cite{ak-s}, is to suitably generalize
Shannon sampling theory for application in the description of Planck scale
physics in curved spacetime. To this end the goal is to find a sampling
theorem that can accommodate variable densities of degrees of freedom,
i.e. varying ``bandwidths" and ``Nyquist rates". In the literature,
several methods for non-equally spaced sampling are known, such as
Gaussian and Lagrangian interpolation, interlaced sampling, or Kramer's,
Papoulis' or Paley-Wiener's sampling theorems, see, e.g.,
\cite{marks,zayed}. These, however, are either specialized or do not
provide a direct handle on the time-varying bandwidth. Indeed, while it is
clear, intuitively, that the bandwidth of a class of signals can be
varying in time, it is surprisingly difficult to give a satisfactory
definition of the concept of time-varying bandwidth:

For example, let us assume a signal $\phi(t)$ obeys a constant bandlimit
$\omega_{max}$. If we consider this signal on some finite interval
$[t_i,t_f]$ and calculate its Fourier series on this interval then one
might expect to find only frequency components smaller or equal to
$\omega_{max}$. This is, however, not the case. Indeed, for any finite
bandwidth one can find signals which oscillate in a given interval
arbitrarily faster than the highest Fourier component. This shows that any
attempt must fail that tries to define a signal's time-varying bandwidth
as the maximal frequency that it contains in a moving temporal window.
There are surprising implications for quantum fields which in this sense
``superoscillate", see \cite{ak-pf-so}.

There is a further reason for why the usual notion of bandwidth does not
lend itself to generalization to obtain a notion of time-variable
bandwidth: the usual definition of bandwidth as the range of the Fourier
spectrum allows to derive Shannon's sampling theorem, but all Fourier
series - they are needed in the proof - are necessarily equidistant.

For this reason, our approach here, as suggested in \cite{ak-s}, is to aim
instead for a generalized sampling theory based on a well-defined notion
of variable Nyquist rate - the minimum rate at which samples must be taken
can be variable. To this end, we will replace the use of Fourier
techniques by the use of more powerful functional analytical techniques.
Sets of sampling points will be spectra of self-adjoint operators: the
spectra of self-adjoint operators do not have to be equidistant.

Technically, our starting point is the observation that if $\phi(t)$ is
bandlimited then the signal $t \phi(t)$ obeys the same bandlimit. This
means that no matter what bandwidth we choose, the corresponding class of
bandlimited signals is invariant under the action of the time operator $T:
~\phi(t) \rightarrow t \phi(t)$. (This is because $T$ is the derivative
operator on frequency space and if a function has support only strictly
within a given interval, then so does its derivative.) The sampling points
will be the spectra of the self-adjoint extensions of the time operator
$T$.
\sn
Technically, the classes of signals with time-varying bandwidth are the
domains of simple symmetric operators $T$ with deficiency indices $(1,1)$
whose self-adjoint extensions possess purely discrete spectra. The
sampling kernel (in the special case of Eq.\ref{e1} this is the term of
the form $\sin(t)/t$) is recognized as representing the matrix elements of
the unitary operators which interpolate the eigenbases of the family of
self-adjoint extensions of $T$. The sampling kernel can be calculated
explicitly in terms of the chosen time-varying Nyquist rate or in terms of
the time-varying ``bandwidth", defined as the inverse of the time-varying
Nyquist rate.
\sn
\section{Sampling theory in Hilbert space}
\subsection{Sampling kernels as unitary operators which
interpolate  self-adjoint extensions} The basic idea in \cite{ak-s} is to
reduce sampling problems to the study of a certain class of operators,
namely the class of simple symmetric operators, $T$, with deficiency
indices $(1,1)$ whose self-adjoint extensions $T(\alpha)$ have purely
discrete spectra $\{t_n(\alpha)\}$. Here, $\alpha \in [0,2\pi ]$ is a real
parameter which parametrizes the $U(1)$-family of self-adjoint extensions
$T(\alpha)$. We denote their eigenvalues and eigenvectors by $t_n(\alpha)$
and $\ve_n(\alpha)$, i.e. \be T(\alpha) \ve_n(\alpha) =
t_n(\alpha)\ve_n(\alpha). \ee As we will see, each such operators possess
the property that the spectra $\{t_n(\alpha)\}$ of its self-adjoint
extensions are covering the real line exactly once. As a consequence,
after fixing the phases of the eigenvectors, each vector ${\bf \phi}$ in
the domain of $T$ is uniquely represented as a function
\be \phi(t)=(\ve,\phi). \ee This immediately provides the connection to
sampling theory:
\mn
Any vector ${\bf \phi}$ is fully determined by its coefficients in one
Hilbert basis, and its coefficients in any arbitrary Hilbert basis can
then be calculated. Here, if a Hilbert space vector's coefficients in the
Hilbert basis $\{\ve_n(\alpha)\}$ of one self-adjoint extension of $T$ are
known, then the coefficient of that vector $\phi$ with any arbitrary
eigenvector ${\bf t}$ of any arbitrary self-adjoint extension of $T$ can
be calculated. Therefore, each function $\phi(t)=( \ve,\phi)$ is already
fully determined for all $t$ if only its values on one of the spectra are
given. For any arbitrary fixed choice of $\alpha$, we obtain the
reconstruction-from-samples formula
 \be \phi(t) = \sum_n
G(t,t_n(\alpha))~\phi(t_n(\alpha)), \label{saab1}
\ee where the sampling kernel $G(t,t_n(\alpha))$ reads \be
G(t,t_n(\alpha))=(\ve,\ve_n(\alpha)). \label{saab2} \ee We have merely
inserted a resolution of the Hilbert space identity. The sampling kernel
is therefore identified as consisting of the matrix elements of the
unitaries which connect the eigenbases of the various self-adjoint
extensions. \sn In order to make this observation into a practical
sampling theorem it will be necessary to establish the relation between
the sampling lattices, i.e. the family of spectra
 $\{t_n(\alpha)\}$, and
the time-varying bandwidth $\omega_{max}(t)$. This will be done in
Sec.\ref{defbw}, using the inverse relation between the bandwidth and the
Nyquist sampling rate. Before, however, let us further clarify the
underlying functional analytic structure.
\subsection{Simple symmetric operators}
We saw that the new and very general sampling theorem follows if we can
show that for each simple symmetric operator $T$ with deficiency indices
$(1,1)$ whose self-adjoint extensions have purely discrete spectra, the
family of spectra $\{t_n(\alpha)\}$ of its self-adjoint extensions
$T(\alpha)$ covers all of $\R$ exactly once.

To this end, let $T$ be a closed symmetric operator on a dense domain
$D_T$ in a complex Hilbert space $H$:
\begin{equation}
(\phi, T\phi) ~\in ~\R  ~~~~~\forall \phi~\in ~D_T,
\end{equation}
Let us choose the scalar product to be antilinear in the first factor.
Recall that $T$ is called simple symmetric, see e.g. \cite{AG}, if it is
not self-adjoint, and if it does not have  any self-adjoint restriction to
an invariant subspace. Note that all symmetric operators are closable.
\sn
As we indicated above, the case of most interest here is that of simple
symmetric operators $T$ which have deficiency indices $(1,1)$, i.e. for
which:
\be
\mbox{dim}\left((T~\pm ~i 1)~D_T\right) ~=~ 1
\ee
It is well known that each such operator has a $U(1)$-family of
self-adjoint extensions $T(\alpha)$, where $\alpha$  labels the
self-adjoint extensions,  e.g. $\exp(i\alpha)\in U(1)$. We consider here
only such $T$ with self-adjoint extensions whose spectra are discrete.
\sn
Note that for finite deficiency indices all self-adjoint extensions have
the same continuous spectrum. Thus, if one of the self-adjoint extension
has a purely discrete spectrum, so do all self-adjoint extensions. As a
consequence, the spectral kernel of each such operator $T$ is empty. This
is because on the one hand any continuous part in its spectral kernel
would also exist in the spectrum of its self-adjoint extensions. On the
other hand, a simple symmetric operator also cannot have any discrete part
in its spectral kernel, since this would contradict its simplicity ($T$
would then have self-adjoint restrictions to eigenspaces). Therefore, for
$T$, each real number $\lambda\in\R$ is a point of regular type, in the
terminology of \cite{AG}, i.e.:
\be
\exists~ k(\lambda)>0~~\forall ~\phi\in D_T:~~~~
\vert\vert (T-\lambda 1)~\phi\vert\vert ~\ge~ k(\lambda)
~\vert\vert\phi\vert\vert
\label{abo}
\ee
It is known that, generally,  if $\lambda$ is a real point of regular type
for a simple symmetric operator $T$ with deficiency indices $(n,n)$, then
there exists a self-adjoint extension $T(\alpha)$ for which $\lambda$ is
an eigenvalue of multiplicity $n$. Thus, $\lambda$ is also an eigenvalue
of $T^*$, of multiplicity at least $n$. On the other hand, if $T$ is any
symmetric operator with finite and  equal deficiency indices $(n,n)$, and
if $\lambda\in \R$ is not in its point spectrum, then the multiplicity of
the eigenvectors of $T^*$ to the eigenvalue $\lambda$ does not exceed the
deficiency index $n$. Thus, if $\lambda$ is a real point of regular type
of the symmetric operator $T$ then the degeneracy of the eigenvalues of
$T^*$ is indeed exactly the deficiency index $n$.
\sn
Here, we are considering operators with deficiency indices $(1,1)$ for
which the entire real line consists of points of regular type. Thus, for
each $t\in \R$ there exists a self-adjoint extension, $T(\alpha)$, for
which $t$ is an eigenvalue. Further, each real number is a nondegenerate
eigenvalue of $T^*$. The overlap of the domain of two self-adjoint
extensions is the domain of $T$. Thus, no eigenvector of $T^*$ can be in
the domains of two self-adjoint extensions.
\sn
We can therefore conclude that for each  simple symmetric operator $T$
with deficiency indices $(1,1)$ whose self-adjoint extensions have purely
discrete spectra,
 the
family of spectra $\{t_n(\alpha)\}$ of its self-adjoint extensions
$T(\alpha)$ covers all of $\R$ exactly once. As we explained above, this
fact directly yields the sampling theorem of Eqs.\ref{saab1},\ref{saab2}.
Note that, since we are working in a complex Hilbert space but wish to
sample real-valued and continuous signals, we will have to prove below
that it is possible to choose the phases of the eigenvectors $\{ \ve\}$
such that the sampling kernel is real and continuous.

\section{Time-varying bandwidths}
\label{defbw}
\subsection{Defining the time-varying bandwidth
as a time-vary\-ing Nyquist rate}  
Different choices of  simple symmetric operators $T$ lead to sampling
formulas for different classes of signals. Let us now identify the
time-varying bandwidths of these classes of signals, as  a function of the
associated family of spectra $\{t_n(\alpha)\}$. As discussed above, it is
impossible to define the notion of time-varying bandwidth in terms of the
signals' Fourier transforms. Our strategy is, therefore, to define a
time-varying bandwidth $\omega_{max}(t)$ directly through a time-varying
Nyquist rate, namely through a time-varying sampling lattice spacing
function $s(t)$. We can then \it define \rm $\omega_{max}(t)=1/2s(t)$. To
this end, we need to show that for the spectra of the self-adjoint
extensions of $T$ the lattice spacing $s(t)$ is indeed a well defined
function of time.

To see this, note that each family of lattices $\{t_n(\alpha)\}$ is
parametrizable such that the $t_n(\alpha)$ are differentiable, strictly
monotonic functions of $\alpha$, obeying, say, $t_n(\alpha +2\pi) =
t_{n+1}(\alpha)$. We will choose the $U(1)$-parametrization as
$e^{i\alpha}$. The $t_n(\alpha)$ are then strictly monotonic since, as we
saw,  each eigenvalue of $T^*$ is nondegenerate and no eigenvector can be
in the domain of two self-adjoint extensions.

Crucially, the strict monotonicity of the $t_n(\alpha)$ implies that, for
a given family $\{t_n(\alpha)\}$, there is exactly one lattice interval
centered around each point $t$. Let us denote the length of this interval
by $s(t)$. Thus, if
\be
t~=~\frac{t_{n+1}(\alpha)+t_n(\alpha)}{2}
\ee
for some $\alpha$, then we set
\be
s(t)~ := ~t_{n+1}(\alpha)-t_n(\alpha)~~.
\ee
For each such operator $T$, the lattice spacing is therefore a
well-defined function of time $s(t)$, i.e. we arrive at a well-defined
time-varying Nyquist rate. Through $s(t)$, we may now define the
time-varying bandwidth as
\be
\omega_{max}(t) := \frac{1}{2s(t)},
\label{bwd}
\ee
Vice versa, given $\omega_{max}(t)$,
the family $\{t_n(\alpha)\}$ can be reconstructed,
up to trivial reparametrization of $\alpha$, through
\be
t_n(\alpha) = t-\frac{1}{4~\omega_{max}(t)}
\ee
and
\be
t_{n+1}(\alpha) = t+\frac{1}{4~\omega_{max}(t)}.
\ee
It is, therefore, equivalent to specify the family of spectra
$\{t_n(\alpha)\}$ or to specify the time-varying Nyquist rate $s(t)$, or
to specify the bandwidth curve $\omega_{max}(t)$.

The conventional characterization of the bandwidth as the width of the
Fourier spectrum becomes applicable only when  the bandwidth is locally
approximately constant. We note that whenever the usual Fourier-based
characterization of the bandwidth is applicable then it yields
$\omega_{max} := 1/2s$ and it therefore agrees with our definition.
\subsection{Bandlimited functions
form the domain of a simple symmetric
time operator $T$}
Technically, we are characterizing a class of bandlimited signals as an
analytic domain of the ``Time operator"  $T: ~T\phi(t)=t\phi(t)$. In this
way we were able to generalize the notion of bandwidth to time-varying
bandwidths. There is a simple argument for why a class of bandlimited
signals should be an analytic domain of $T$ (i.e. a domain on which $T$
can be applied an arbitrarily finitely many times). To this end, we need
to begin with precise definitions in the case of constant bandwidth of
what we mean by  bandlimited functions and what we mean by strictly
bandlimited functions:
\sn
We define the class $\B$ of \it strictly bandlimited \rm signals with
bandwidth $\omega_{max}$ as consisting of all those signals whose Fourier
transform $\tilde{\phi}(\omega)$ has the properties that all its
derivatives $\partial^n_\omega\tphi(\omega)$  exist and are square
integrable for all $n$, and that there exists a
$\omega_0(\phi)<\omega_{max}$ so that
\be
\tilde{\phi}(\omega)=0 ~~~~~\mbox{for all}
 ~~\vert\omega\vert > \omega_0(\phi)
\label{sbl}
\ee
We define the scalar product of two signals through
\be
(\phi_1,\phi_2) ~=~ \int_{-\omega_{max}}^{\omega_{max}}
d\omega~\tphi_1(\omega)^* \tphi_2(\omega)
\label{sppp}
\ee
The closure of $\B$ with respect to the norm induced by Eq.\ref{sppp}
yields the Hilbert space $\Hi:=\overline{\B}$ which we call the space of
all \it bandlimited signals. \rm (What we really mean is of course a space
of physical fields that obey an ultraviolet cutoff).
\sn
Now we observe that if a signal $\phi(t)$ is  strictly bandlimited, i.e.
obeys Eq.\ref{sbl}, then also the signal $T\phi(t)=t\phi(t)$ is strictly
bandlimited and obeys Eq.\ref{sbl}.
 This is because
the Fourier transform of the multiplication operator $T:~~T\phi(t) =
t\phi(t)$ is the differentiation operator. Clearly the action of the
differentiation operator action changes the support of the spectrum at
most infinitesimally. Thus, if the signals Fourier transform has support
strictly only inside the bandwidth  interval, then also the derivative of
the signals Fourier transform has support strictly only within the
bandwidth interval.
\sn
Crucially, the fact that multiplying a signal $\phi(t)$ with the time
variable, $\phi(t)\rightarrow t\phi(t)$, affects the bandwidth of the
signal only infinitesimally, is \it independent \rm of  which bandwidth
has been chosen. This means that, independently of the chosen bandwidth,
the strictly bandlimited signals form an analytic domain of the time
operator.
\sn
For this reason it is clear that any proper definition of time-varying
bandwidths should have the property that each class of signals with
time-varying bandwidth forms an analytic domain of the time operator.
\sn
Finally, we remark that, technically, we obtain $D_T$ from $\B$ by closing
the operator (self-adjoint extensions are most conveniently constructed
when starting from an operator which is closed), and we obtain $\Hi$ by
closing the domain $D_T$ (or $\B$).

\subsection{The bandwidth limits its own variability}
Intuitively, consistency conditions
on the possible bandwidth curves $\omega_{max}(t)$
are to be expected.
This is because the bandwidth
 $\omega_{max}(t)$ poses a
limit on how much signals can vary around time
$t$ and the bandwidth should therefore also pose a limit on how much it,
itself,  the bandwidth $\omega_{max}(t)$, can vary around time $t$.
Indeed, there already follows a simple consistency
condition from its definition in Eq.\ref{bwd}:
\sn
 Starting from
\be
\frac{d}{dt} \omega_{max}(t) ~=~ -\frac{1}{2~s^2(t)}~\frac{ds(t)}{dt}
\ee
and using
\be
\frac{ds(t)}{dt} ~=~\frac{ds(\alpha)}{d\alpha}~\frac{d\alpha(t)}{dt}
\ee
we calculate separately
\be
\frac{ds(\alpha)}{d\alpha}~=~\frac{d}{d\alpha}
\left(t_{n+1}(\alpha)-t_n(\alpha)\right)
\ee
and
\be
\frac{d\alpha(t)}{dt}~=~\left(\frac{d}{d\alpha}~
\frac{t_n(\alpha)+t_{n+1}(\alpha)}{2}\right)^{-1}.
\ee
Thus, introducing the notation
\be
t_r^\prime(\alpha) := \frac{d}{d\alpha}~t_r(\alpha)
\ee
we obtain:
\be
\frac{d}{dt}\omega_{max}(t)   ~=~ -4 ~\omega^2_{max}(t)
~\frac{t_{n+1}^\prime(\alpha)-
t_n^\prime(\alpha)}{t^\prime_{n+1}(\alpha)+t^\prime_n(\alpha)}
\label{oucr}
\ee
Eq.\ref{oucr} implies, in particular, a limit which the bandwidth
$\omega_{max}(t)$ imposes on how much $\omega_{max}(t)$ itself can vary:
\be
\vert~ d\omega_{max}(t)/dt~\vert   ~<~ 4 ~\omega^2_{max}(t)
\ee
\subsection{The (varying) bandwidth limits how much a bandlimited signal can
be peaked}
\label{five} We will now show that if a signal $\phi(t)$ is strictly
bandlimited by a constant $\omega_{max}$, and if $\phi(t)$ is sampled at
the Nyquist rate, then the standard deviation $\Delta T$ of the sampled
values is at least $\Delta T\ge 1/4\omega_{max}$. Thus, the smaller the
bandwidth, the larger is the signals' minimum standard deviation, i.e. the
less they can be peaked. A time-varying bandwidth means that the lower
bound to the standard deviation of the signals' samples depends on the
time $t$ around which the signal is centered (its first moment).

We saw that for each Nyquist rate there is a whole family of sampling
lattices. We will find that the standard deviation (as well as all other
moments) of the sampled values of a  strictly bandlimited signal do not
depend on the lattice on which the samples are taken.

\subsubsection*{The minimum standard deviation $\sigma_{min}(t)$}
We have defined time-varying bandwidths as time-varying Nyquist rates. Let
us now investigate in which sense such a time-varying bandwidth is
limiting how much signals can vary over time. We will find that the
time-varying Nyquist rate is limiting how much the signals can be peaked
at different times.
\sn
Consider for example a normalized
signal $\phi(t)$
\be
(\phi,\phi)~=~1
\ee
which is centered around the time $t$ in the sense
that $t$ is the expectation value of the time
operator $T$, i.e. its first moment:
\be
(\phi, T \phi) ~=~ t
\ee
A measure of how much the signal $\phi(t)$
is peaked around its expectation
value is for example its second moment,
i.e. its formal standard deviation:
\be
\Delta T (\phi)~ = ~ \sqrt{(\phi,T^2 \phi)-(\phi,T\phi)^2}
\label{sdev}
\ee
In order to explicitly evaluate Eq.\ref{sdev} we can
insert any one of the resolutions of the identity
$1=\sum_n \ve^*_n\otimes \ve_n$ induced
by the self-adjoint extensions $T(\alpha)$
of $T$, to obtain:
\be
\left( \Delta T(\phi)\right)^2 ~=~ \sum_n
\phi(t_n(\alpha))^2 ~ t_n^2(\alpha) ~-~ \left(
\sum_m\phi(t_m(\alpha))^2~ t_m(\alpha)                     \right)^2
\label{evalu}
\ee
Thus, $\Delta T(\phi)$ is the standard deviation of the discrete set of
samples of the signal $\phi(t)$ on a sampling lattice $\{t_n(\alpha)\}$.
The standard deviation of a signal's samples $\phi(t_n(\alpha))$ on the
lattice $\{t_n(\alpha)\}$ is clearly the same as the standard deviation of
the signal's samples $\phi(t_n(\alpha^\prime))$ on any other lattice
$\{t_n(\alpha^\prime)\}$ of the family.

We note, that therefore in the  Hilbert space of bandlimited signals, the
standard deviation $\Delta T(\phi)$ is not the usual standard deviation of
a continuous function, which would be calculated in terms of integrals,
but instead it is the standard deviation of the discrete set of samples of
the signal, taken on a sampling lattice.
\mn
Our claim is that for the class of signals which is characterized by some
time-varying Nyquist rate $s(t)$ there is a function $\sigma_{min}(t)$
which sets a time-varying finite lower bound to the standard deviation
$\Delta T$ of the samples of signals which are centered around $t$:
\be
\Delta T(\phi) ~\ge ~\sigma_{min}(t) ~~~~~~~\mbox{for all }~~\phi\in D_T
~~\mbox{which obey}~ ~(\phi,T\phi)=t
\ee
Concretely, this means that if a normalized
signal is sampled on a sampling lattice $\{t_n(\alpha)\}$
(where $\alpha$ is arbitrary) and
if the average, or first moment, of the samples is $t=
 \sum_n \phi(t_n(\alpha))^2~ t_n(\alpha)$ then we claim that
the standard deviation of the samples
is larger or equal than a finite value $\sigma_{min}(t)$:
\be
\sqrt{\sum_n \phi(t_n(\alpha))^2 ~ t_n^2(\alpha)
~-~ \left(\sum_m\phi(t_m(\alpha))^2~ t_m(\alpha) \right)^2}
\ge \sigma_{min}(t)
\ee
Let us calculate the minimum standard deviation $\sigma_{min}(t)$ from the
Nyquist rate $s(t)$, i.e. from the bandwidth $\omega_{max}(t)=1/2s(t)$.

\subsubsection*{Calculating $\sigma_{min}(t)$}
In order to calculate $\sigma_{min}(t)$ for strictly bandlimited signals,
it will be convenient to calculate first the minimum standard deviation
curve $\sigma^{(\alpha)}_{min}(t)$ for a larger set signals, namely all
those in the domain $D_{T(\alpha)}$ of a self-adjoint extension
$T(\alpha)$ of $T$. (We recall that the dense set of strictly bandlimited
signals is an analytic domain of $T$, i.e. they are contained in $D_T$).
\sn
In the domain $D_{T(\alpha)}$ of a self-adjoint extension $T(\alpha)$,
 there are eigenvectors $\ve_n(\alpha)$ of
$T(\alpha)$ and clearly each is centered around time $t_n(\alpha)$ and has
vanishing standard deviation. We recall that the standard deviation
vanishes exactly only for eigenvectors. Thus, in the domain
$D_{T(\alpha)}$ of a given self-adjoint extension, the minimum standard
deviation vanishes at the eigenvalues
\be
\sigma^{(\alpha)}_{min}(t_n(\alpha))~=~0.
\ee
On the other hand, for all vectors in $D_{T(\alpha)}$
 which are
centered around a point in time $\tau$ which is in between two neighboring
eigenvalues, say $t_n(\alpha) < \tau < t_{n+1}(\alpha)$ the standard
deviation cannot vanish, i.e. we expect that:
 \be
\sigma^{(\alpha)}_{min}(\tau))\neq 0.
\ee
In order to calculate the minimum standard deviation,
$\sigma^{(\alpha)}_{min}(\tau)$,
we use that  the most peaked signal  $\phi_\tau \in D_{T(\alpha)}$ which is
centered around $\tau$
must be a linear combination of the closest two eigenvectors, namely
\be
\phi_\tau ~=~a_1 \ve_n(\alpha) + a_2 \ve_{n+1}(\alpha)
\ee
The requirements $(\phi_\tau,T\phi_\tau) =\tau$ and
$(\phi_\tau,\phi_\tau)=1$ determine the coefficients:
$a_1=(t_{n+1}-\tau)/(t_{n+1}-t_n)$, and $a_2 = (\tau-t_n)/(t_{n+1}-t_n)$.
The standard deviation $\Delta T(\phi_\tau)$ of the signal $\phi_\tau$
which is maximally peaked around the time $\tau$ is then easily calculated
and yields the minimum standard deviation $\sigma^{(\alpha)}_{min}(\tau)$:
\begin{eqnarray}
\sigma^{(\alpha)}_{min}(\tau) & = & \Delta T(\phi_\tau)\nonumber \\
 \\ & = &  \sqrt{(\phi_\tau,T^2\phi_\tau)-\tau^2} \\
\label{hc7}\nonumber  \\
  & = &
\sqrt{(\tau -t_n) (t_{n+1}-\tau)} \nonumber
\end{eqnarray}
It is clear that for each choice of $\alpha$ the minimum standard
deviation $\sigma^{(\alpha)}_{min}$ is zero at all times
$\{t_n(\alpha)\}$, since these are the eigenvalues of $T(\alpha)$.
Eq.\ref{hc7} now shows that in the times between any two neighboring
eigenvalues $t_n(\alpha)$ and $t_{n+1}(\alpha)$ the minimum standard
deviation curve $\sigma^{(\alpha)}_{min}(t)$ is simply a half circle with
diameter equal to the lattice spacing $s=t_{n+1}(\alpha)-t_n(\alpha)$.
\sn
We can now calculate
the minimum standard deviation curve $\sigma_{min}(t)$ for the
for the vectors in $D_T$, from the knowledge
of the minimum standard deviation curves $\sigma^{(\alpha)}_{min}(t)$
of the self-adjoint extensions $T(\alpha)$.

This is because $D_T\subset D_{T(\alpha)}$ for all $\alpha$, and therefore
the minimum standard deviation curve $\sigma_{min}(t)$ for the vectors in
the domain $D_T$  is bounded from below by all the minimum standard
deviation curves $\sigma^{(\alpha)}_{min}(t)$ of the self-adjoint
extensions. We obtain:
\be
\sigma_{min}(t) ~=~\max_{{}_{ \alpha \in U(1)}}\sigma^{(\alpha)}_{min}(t)
\label{sigmamin}
\ee
Since strictly bandlimited signals are elements of $D_T$, this means that
strictly bandlimited signals centered around time $t$ are peaked at most
as much as specified by the minimum standard deviation $\sigma_{min}(t)$.
Here, $\sigma_{min}(t)$ is the standard deviation of the discrete samples
of the signal on (any arbitrary) one of the family of sampling lattices
$\{t_n(\alpha)\}$ with the Nyquist rate $s(t)$.

In physical terms, this means that a finite ultraviolet cutoff of this
kind, constant or varying, corresponds one-to-one to the existence of a
constant or varying finite minimum to the formal uncertainty in position
(or time), as mentioned above in the context of studies in quantum gravity
and string theory, see \cite{qg}.

\subsubsection*{$\sigma_{min}$ in the case of constant bandwidth}
In the case of a constant bandwidth, Eq.\ref{sigmamin} is readily evaluated:
\sn
For a constant bandwidth $\omega_{max}$, the lattice spacing is
$s=1/2\omega_{max}$. Thus, each curve $\sigma^{(\alpha)}_{min}(t)$
consists of half circles with radius $1/4\omega_{max}$ which stretch from
eigenvalue to eigenvalue. Therefore, from Eq.\ref{sigmamin}, the minimum
standard deviation curve $\sigma_{min}(t)$ for strictly bandlimited
signals is a constant:
\be
\sigma_{min}(t) ~=~ \frac{1}{4\omega_{max}}
\label{mindiv}
\ee
Let us confirm this result by explicitly solving the variational problem
 for finding the minimum standard deviation:
\sn
Our aim is to
minimize $(\Delta T)^2$ for $\phi \in D_T$
by minimizing $(\phi, T^2 \phi)$ while enforcing the constraints
$(\phi,\phi)=1$ and $(\phi, T\phi)=t$.
\sn
To this end, we represent $T$ in Fourier space as $T= -i ~d/d\omega$ on
strictly bandlimited signals, i.e. on signals which in Fourier space obey,
in particular, the boundary condition:
\be
\tilde{\phi}(-\omega_{max}) ~=~ 0~=~
\tilde{\phi}(\omega_{max})
\ee
Introducing Lagrange multipliers $k_1,k_2$, the functional to be minimized
is, therefore
$$ $$
\be
S=\int_{-\omega_{max}}^{\omega_{max}} d\omega
~\left\{(\partial_\omega{\tilde{\phi}}^*)(\partial_\omega{\tilde{\phi}}) +
k_1 ({\tilde{\phi}}^*{\tilde{\phi}} - c_1) + k_2 (-i
{\tilde{\phi}}^*\partial_\omega{\tilde{\phi}}-c_2)\right\},
\ee
$$ $$
which yields the Euler-Lagrange equation:
\be
-\partial_\omega^2 {\tilde{\phi}} + k_1 {\tilde{\phi}} -i k_2
\partial_\omega{\tilde{\phi}} ~=~0
\ee
For each value of a $T$-expectation value $t$, there exists exactly one
(up to phase) normalized solution which also obeys the boundary condition,
namely:
\be
{\tilde{\phi}}_t(\omega)~=~ \frac{1}{\sqrt{2\pi \omega_{max}}}
~\cos\left(\frac{\pi~ \omega}{2 ~\omega_{max}}\right)~e^{2 \pi  i t \omega}
\ee
The standard deviations $\Delta T(\phi_t)$ of these solutions are
straightforward to calculate in Fourier space, to obtain
\be
\Delta T(\phi_t) ~=~ \frac{1}{4\omega_{max}}~~~~~\mbox{for all $t$}
\ee
which indeed coincides with our previous result,
Eq.\ref{mindiv}, for the minimum standard deviation
$\sigma_{min}$ in the class
of strictly bandlimited signals of constant bandwidth $\omega_{max}$.

\subsubsection*{Relation to superoscillations}
For the special case when the bandwidth is constant, the property of
strictly bandlimited signals to obey a minimum standard deviation may
appear to be  not more and not less intuitive than the conventional belief
that bandlimited signals are slow-varying, namely that bandlimited signals
are varying at most as fast as their highest Fourier component.

Let us emphasize, however, that the conventional belief that bandlimited
signals cannot vary faster than their fastest Fourier component does not
hold true. This has been pointed out in particular in
\cite{aharonovetal,berry}, where counterexamples have been given and named
``superoscillations". In \cite{berry}, Berry quotes I. Daubechi in
explaining that phenomena related to superoscillations have also been
known to occur as instabilities in oversampling. For related work, see
also e.g. \cite{berry2}.

On the other hand, our finding that strictly bandlimited signals obey a
minimum standard deviation is true for all strictly bandlimited signals,
without exceptions. Superoscillations are naturally accommodated in this
framework in the sense that strictly bandlimited signals can contain
arbitrarily sharp spikes, and that they can therefore be locally varying
arbitrarily faster than their highest Fourier component, while still
obeying the standard deviation constraint, see \cite{ak-so}.

\section{Types of sampling lattices} 

\subsection{Unbounded and semibounded sampling lattices} 
In those cases where the lattices of sampling times are ranging from
$-\infty$ to $+\infty$, the operator $T$ and its self-adjoint extensions
are fully unbounded operators.

Since simple symmetric operators can be semibounded, it is also possible
to choose sampling lattices which are merely semibounded, i.e. which range
  from $-\infty$ to some finite largest sampling time $t_N$, or of course
  from some finite sampling time to $+\infty$. However, we found above, on
functional analytical grounds, that the family of spectra always covers
all of $\R$, i.e. the sampling formula always allows to reconstruct
$\phi(t)$ for all $t\in \R$. It appears to be paradoxical that signals can
be sampled on a semibounded lattice but recovered for all $t\in \R$.
Nevertheless, it is a fact that simple symmetric operators can be
semi-bounded and that in such cases $\phi(t)$ can still be reconstructed
for all $t\in \R$ from its values sampled on a semibounded lattice.

The resolution to the paradox is that the bandwidth curve
$\omega_{max}(t)$ rapidly decays after the last sampling point $t_N$. This
means that the class of signals which are sampled in such cases is a class
of signals which quickly become constant in the interval from $t_N$ to
infinity.

To see this, assume that the spectrum of the self-adjoint extension $T(0)$
is, say, bounded from above by possessing a largest eigenvalue $t_N$. We
recall that each real number appears as an eigenvalue in one of the
self-adjoint extensions of $T$. This means that as $\alpha$ runs through
the family of self-adjoint extensions, the highest eigenvalue
$t_N(\alpha)$ of their respective spectra moves from $t_N(0)=t_N$ to
infinity, $t_N(\alpha) \rightarrow +\infty$ for $\alpha \rightarrow 2\pi$,
while the next to highest eigenvalue behaves as
$t_{N-1}(\alpha)\rightarrow t_N$.

Thus, while $\alpha$ runs through the family of self-adjoint extensions,
the center $t=(t_N(\alpha)-t_{N-1}(\alpha))/2$ of the interval between the
last two eigenvalues moves to infinity, and thereby the spacing $s(t)=
t_N(\alpha)-t_{N-1}(\alpha)$ of this interval diverges. Thus, the
bandwidth $\omega_{max}(t) =1/2s(t)$ indeed decays like $\omega_{max}(t)
~\rightarrow ~ 1/t$ between the last point of the sampling lattice and
infinity.

We will show in Sec.\ref{effi} that this phenomena can be used to
sample signals on lattices which are not actually finite lattices, but which are
effectively finite lattices for all practical purposes.

\subsection{No finite sampling lattices} 
\label{nofin} Before we discuss sampling on effectively finite sampling
lattices, let us convince ourselves that there are no actually finite
lattices of sampling times:

To this end, we recall that we are defining the class of signals to be sampled as
the representation of the multiplication operator $T:~~T\phi(t)=t\phi(t)$
on the space of functions for which the scalar product
\begin{equation}
( \phi_1,\phi_2) :=
\sum_n \phi^*_1(t_n(\alpha))~ \phi_2(t_n(\alpha))
\label{spr}
\end{equation}
is independent of $\alpha$, where $\{t_n(\alpha)\}$ are the
spectra of a family of self-adjoint extensions.
\sn
For example, the
Shannon sampling formula is now the expansion of signals
\be
\phi(t) = \sum_n c^{(\alpha)}_n~b^{(\alpha)}_n(t)
\ee
with coefficients
\be c^{(\alpha)}_n = \phi(t_n(\alpha))
\ee
 in the Hilbert basis of functions:
\be
b^{(\alpha)}_n(t)= \frac{\sin[2\pi~ (t-t_n(\alpha))~\omega_{max}]}{
2\pi~ (t-t_n(\alpha))~\omega_{max}}
\ee
The basis functions $b^{(\alpha)}_n(t)$ are orthogonal with respect to the
scalar product given by Eq.\ref{spr}. Here, $\{t_n(\alpha)\}$ is the
family of  lattices with the equidistant lattice points defined as in
Eq.\ref{sisp} through $t_n(\alpha) = \alpha + n/2\omega_{max}$. We recall
that for vectors in the Hilbert space, the scalar product Eq.\ref{spr}
yields the same result for any arbitrary choice of  $\alpha$.
\sn
Here, and also in the case of generic nonequidistant
sampling lattices $\{t_n(\alpha)\}$, the $\alpha$-independence of
the scalar product implies the existence of isometries between the
$l^2$-Hilbert spaces over each of the lattices
of the family $\{t_n(\alpha)\}$.
The matrix elements of these isometries then form the sampling kernel $G$.
\sn
This means that the signals are solutions to the equation
\begin{equation}
\phi(t_m({\alpha^\prime}))~ = ~\sum_n
~G(t_m({\alpha^\prime}),t_n(\alpha))~\phi(t_n(\alpha)),
\label{s1}
\end{equation}
for all $\alpha$ and $\alpha^\prime$.
\sn
In addition, the signals ${\bf \phi}$ form a representation of
$T$, i.e. they also obey
\begin{equation}
t_m({\alpha^\prime})~\phi(t_m({\alpha^\prime})) = \sum_n
~G(t_m({\alpha^\prime}),t_n(\alpha))~t_n(\alpha)~\phi(t_n(\alpha))
\label{s2}
\end{equation}
for all $\alpha$ and $\alpha^\prime$.
Assume now that the family $\{t_n(\alpha)\}$ consists of
  finite lattices, say with $N$ lattice sites.
The sampling kernel $G$ is then a
$N \times N$- dimensional matrix with matrix elements
$G_{mn}(\alpha^\prime,\alpha):=G(t_m(\alpha^\prime),t_n(\alpha))$.
In this case, Eqs.\ref{s1},\ref{s2}
would yield the following equation for  the
sampling kernel matrix $G=G_{nm}(\alpha^\prime,\alpha)$:
\sn
\begin{equation}
\left(
\matrix{ t_1(\alpha^\prime)
 & \quad &  \mbox{\Large 0}  \cr
                  \quad &   \ddots & \quad \cr
                \mbox{\Large 0} &        \quad & t_{N}({\alpha^\prime})  \cr
} \right)
~~=~~ G ~
\left(
\matrix{ t_1(\alpha)
 & \quad &  \mbox{\Large 0}  \cr
                  \quad &   \ddots & \quad \cr
                  \mbox{\Large 0} &        \quad & t_{N}({\alpha})  \cr
} \right) ~
G^{-1}
\end{equation}
$$ $$
It is clear that this equation cannot be solved in finite dimensions,
since the adjoint action of any isometric matrix $G$ would be spectrum
preserving. This simply reflects the functional analytic fact that finite
dimensional symmetric operators are automatically self-adjoint, i.e. that
only unbounded operators can be simple symmetric.

\subsection{Effectively finite sampling lattices}
\label{effi} In practical applications in engineering it is only possible
to use finite lattices to sample actual signals, i.e. the theoretically
infinite lattices are cut off.
\sn
The generalized sampling theorem allows us to perform the cutoff of
lattices in a simple and controlled way. Namely, outside the finite time
span, say $[t_{lo},t_{up}]$ in which actual sampling takes place, we are
free to choose arbitrarily large lattice spacings. By choosing very large
lattice spacings outside the interval $[t_{lo},t_{up}]$ we achieve that
the bandwidth quickly decays to zero outside the interval. More precisely,
the bandwidth decays like $1/\vert t-t_{up}\vert$ and $1/\vert
t-t_{lo}\vert$ to the left and right of the boundaries of the interval
$[t_{lo},t_{up}]$, meaning that we are sampling signals which quickly
become constant outside the interval. Since the signals are in the Hilbert
space and are therefore square summable, they indeed decay to the constant
zero. This shows that while the unboundedness of the operator $T$ is a
mathematical necessity this does not affect in any essential way the
sampling properties of the signals (or fields) that are based on finite
sets of samples. After we develop the details of the sampling theorem we
will come back to effectively finite sampling lattices in
Sec.\ref{secsummary}.

\section{Calculating the sampling kernel}
\subsection{Embedding a self-adjoint operator into a $U(1)-$
 family of self-adjoint operators}
\label{calsk}
We begin by specifying a lattice of sampling times
$\{t_n\}$ on which signals are to be sampled and we define the self-adjoint
operator $T(0)$ which has this lattice as its spectrum.

Our strategy is to
define a self-adjoint operator $T(0)$
with the lattice  $\{t_n\}$ as its spectrum,
then to use the data $\{t_n^\prime\}$
to specify a simple symmetric restriction $T$
of $T(0)$, and finally to construct its $U(1)$- family of
self-adjoint extensions $T(\alpha)$.
The scalar product of their eigenvectors
yields the desired sampling kernel $G(t,t_n)$.
\smallskip\newline
To this end, consider the self-adjoint operator
\begin{equation}
T(0) := \sum_n t_n ~{\ve_n}\otimes{\ve_n^*} \end{equation}
on its domain $D_{T(0)}$. Here, $\ve_n$ denotes the eigenvector to
the eigenvalue $t_n$, and $\bf \ve_n^* \rm $ denotes the dual vector.
The Cayley transform of $T(0)$ is a unitary operator
\begin{equation}
U(0) = \sum_n ~\frac{t_n-i}{t_n+i}~ \ve_n\otimes {\ve_n^*}
\end{equation}
which is defined on the entire Hilbert space $H=\bar{D}_{T(0)}$.
We restrict $U(0)$ to a simple isometric operator
\begin{equation}
S= U(0)\vert_{H \ominus C {\bf v_+}}
\end{equation}
by taking from its domain one dimension spanned by a
normalized vector
 ${\bf v_+} \in H$.
The inverse Cayley transform then yields a
 simple symmetric operator $T$ on $D_T$
\begin{equation}
D_T = \left\{ {\bf \phi} \in D_{T(0)}
~\vert ~ \left( {\bf v_+}, (T(0) +i)
{\bf \phi}\right) = 0 \right\}
\end{equation}
iff the so-defined $D_T$ is dense and contains no eigenvectors.
In the $\{{\ve_n}\}$-basis, the condition
$\left( {\bf v_+},  (T(0) +i) {\bf \phi}\right) = 0$ reads
$\sum_n( {\bf v_+},{\ve_n})~ (t_n+i)~(
{\ve_n}, \phi) =0$. Thus,
for $D_T$ to be dense, ${\bf v_+}$
must be chosen such that
\begin{equation}
\sum_n \vert( {\bf v_+},{\ve_n})~(t_n +i)\vert^2~=~\mbox{divergent}
\end{equation}
since otherwise all of $D_T$ would be orthogonal  to
$\bf w \rm := \sum_n (t_n+i) ~({\bf v_+} , \ve_n)~\ve_n$,
which would then be a vector in $H$.

Further, the condition that there are
no eigenvectors in $D_T$ implies that ${\bf v_+}$
must be chosen such that $( {\bf v_+},
t_n) \neq 0$ for all $n$.
The second deficiency space is of course spanned
by ${\bf v_-} = U(0) ~{\bf v_+}$.
Now a $U(1)$-family of self-adjoint extensions $T(\alpha)$
of $T$ is obtained as the inverse Cayley transforms
of the unitary extensions $U(\alpha)$ of $S$:
\begin{equation}
U(\alpha) = \left(1 - {\bf v_-}\otimes {\bf v_-^*}
+ e^{i\alpha}~{\bf v_-}\otimes
{\bf v_-^*} \right) U(0)
\end{equation}
The choice of a spectrum $\{t_n\}=\{t_n(0)\}$ and
a deficiency vector ${\bf v_+}$ thus determines
a family of self-adjoints $T(\alpha)$ and unitaries $U(\alpha)$
with their eigenbases  $\{\ve_n(\alpha)\}$ (up to phases).
The functions $t_n(\alpha)$ are strictly monotonic since
each eigenvalue of $T^*$ is nondegenerate
and no eigenvector can be in the domain
of two self-adjoint extensions.
\subsection{Calculating the $t$-dependence of the
sampling kernel}
The aim is to calculate the sampling kernel, i.e. the
scalar products of the eigenvectors of the unitary and
self-adjoint extensions. To this end, let
${\ve_r(\alpha)}$ be a normalized
eigenvector of the operators $U(\alpha)$ and $T(\alpha)$. Then:
\begin{eqnarray*}
\frac{t_r(\alpha)-i}{t_r(\alpha)+i}~~
\left(\ve_n,\ve_r(\alpha)\right) & = &
({\ve_n}, U(\alpha)
{\ve_r(\alpha)}) \\
& = & \left({\ve_n}, (1+
 (e^{i\alpha}-1) ~{\bf v_-}\otimes {\bf v_-^*}
)~U(0) ~{\ve_r(\alpha)}\right)\\
& = & \frac{t_n -i}{t_n+i}~\left(
\ve_n, {\ve_r(\alpha)}\right) + (e^{i\alpha}-1)
\left({\ve_n}, {\bf v_-}\right) \left({\bf v_+},{\ve_r(\alpha)}\right)
\end{eqnarray*}
Thus,
\begin{equation}
\left({\ve_n}, {\ve_r(\alpha)}\right)~ =~  \frac{
f_n ~ N(t_r(\alpha))
}{t_r(\alpha)- t_n}
\label{samker}
\end{equation}
where we defined
\begin{equation}
N(t_r(\alpha)) := \frac{t_r(\alpha)+i}{2i}~(
 e^{i\alpha}-1)~\left( {\bf v_+}, {\ve_r(\alpha)}\right)
\label{e19}
\end{equation}
and
\be
f_n := (t_n+i)~\left({\ve_n}, {\bf v_-}\right)
\label{deff}
\ee
In this way we have separated the $t_r(\alpha)$-dependence of the sampling
kernel in the form of $N(t_r(\alpha))$ from its $n$-dependence through the
coefficients $f_n$. It is now straightforward to calculate
$N(t_r(\alpha))$, i.e. the $t_r(\alpha)$-dependence of the sampling kernel
by using the normalization condition:
\begin{equation}
1 = \sum_n\left( \ve_r(\alpha),{\ve_n}\right)~
\left({\ve_n}, {\ve_r(\alpha)}\right) = \sum_n~
\frac{ f_n^*~f_n~N^*(t_r(\alpha))N(t_r(\alpha))}{(t_r(\alpha)-t_n)^2}
\end{equation}
Thus, for all $t$ in the family $\{t_r(\alpha)\}$,
i.e. for all $t\in \R$,  there holds
\begin{equation}
\vert N(t)\vert ~ =~ \left( \sum_n\frac{f_n^*f_n}{(t-t_n)^2}\right)^{-1/2}
\label{Ne}
\end{equation}
\subsection{Calculating the $n$-dependence of the
 sampling kernel}
In order to calculate the coefficients $f_n$, we begin by noting that
\begin{eqnarray*}
\left({\ve_n}, {\bf v_-}\right) & = & \left({\ve_n}, U(0){\bf v_+}\right)\\ \\
 & = & \sum_m~\left({\ve_n},U(0)\ve_m\right)\left(\ve_m,{\bf v_+}\right) \\ \\
 & = & \sum_m~\left({\ve_n}, \ve_m\right)~
 \frac{t_m-i}{t_m+i}~\left( \ve_m,{\bf v_+}\right)
 \\ \\
 & = & \frac{t_n-i}{t_n+i}~\left({\ve_n}, {\bf v_+}\right)
\end{eqnarray*}
so that, in addition to Eq.\ref{deff}, the $f_n$ also obey
\be
f_n~=~(t_n-i)~\left({\ve_n}, {\bf v_+}\right)
\label{vp}
\ee
We will now show that $t_n^\prime \propto f_n^*f_n $, i.e., more precisely,  that
$dt_n(\alpha)n/d\alpha\vert_{\alpha =0}=f_n^*f_n/2$.

To this end, inserting a resolution of the identity
 in terms of the $\{\ve_n\}$-eigenbasis into Eq.\ref{e19} yields
\be
N(t_r(\alpha))
  =  \frac{t_r(\alpha) +i}{2i} (e^{i\alpha}-1)
\sum_n~\left( {\bf v_+},
 \ve_n\right)~\left({\ve_n}, {\ve_r(\alpha)}\right)
\ee
Inserting Eq.\ref{samker} and Eq.\ref{vp}, we obtain
\be
N(t_r(\alpha))  = \frac{t_r(\alpha) +i}{2i} (e^{i\alpha}-1)
 \sum_n \frac{f_n^*f_n ~~N(t_r(\alpha))}{(t_n+i)
(t_r(\alpha) -t_n)}
\ee
Thus:
\begin{equation}
\frac{2i}{e^{i\alpha}-1} ~=~
\sum_n  \frac{f_n^* f_n~~(t_r(\alpha)+i)}{(t_n+i)(t_r(\alpha)-t_n)}
\label{ia}
\end{equation}
Let us calculate separately the imaginary and the real part of Eq.\ref{ia}.
The imaginary part yields:
\be
\frac{1}{e^{i\alpha} -1} + \frac{1}{e^{-i\alpha}-1}~=~
\frac{1}{2i} ~\sum_n~ f_n^* f_n~
~\frac{(t(\alpha)+i)(t_n-i)-(t(\alpha)-i)(t_n+i)}{(t_n^2+1)(t(\alpha)-t_n)}
\ee
and thus,
\be
-1 = \sum_n \frac{-f_n^* f_n}{t_n^2+1}
\label{normvp}
\ee
Using Eq.\ref{vp}, we see that Eq.\ref{normvp} simply expresses the
normalization condition for ${\bf v_+}$.
\sn
On the other hand, the real part of Eq.\ref{ia} yields:
\be
\frac{1}{e^{i\alpha}-1} - \frac{1}{e^{-i\alpha}-1} ~=~ \frac{1}{2i}~
\sum_n f_n^* f_n~\left(\frac{ t(\alpha) +i}{(t_n+i)(t(\alpha)-t_n)}
+\frac{t(\alpha-i}{(t_n-i)(t(\alpha)-t_n)}\right),
\ee
$$ $$
\be
i~\frac{e^{-i\alpha} - e^{i\alpha}}{2-e^{i\alpha}-e^{-i\alpha}}~=~
\frac{1}{2} \sum_nf_n^* f_n \frac{(t(\alpha)+
i)(t_n-i)+(t(\alpha)-i)(t_n+i)}{
(t_n^2+1)(t(\alpha)-t_n)},
\ee
and thus, using
\begin{eqnarray*}
\cot(\alpha/2)  & = & ~ i~\frac{e^{i\alpha/2}+
e^{-i\alpha/2}}{e^{i\alpha/2}-e^{-i\alpha/2}}\\
 & = &  i~\frac{e^{-i\alpha} - e^{i\alpha}}{2-e^{i\alpha}-e^{-i\alpha}}
\end{eqnarray*}
we obtain:
\begin{equation}
\cot\left(\alpha/2\right) = \sum_n \frac{f_n^* f_n~~
(t_n t_r(\alpha) +1)}{(t_n^2 +1)(t_r(\alpha)-t_n)}
\label{co}
\end{equation}
Since $t_r(\alpha)\rightarrow t_r$ for
 $\alpha\rightarrow 0$,  Eq.\ref{co}
finally yields:
\be
\frac{dt_r(\alpha)}{d\alpha}\vert_{\alpha =0}~=~\frac{f_r^*f_r}{2}
\label{fbetr}
\ee
\subsection{The deficiency vector ${\bf v_+}$ is
parametrized by the coefficients $f_n$}
Through Eq.\ref{vp},
the deficiency vector ${\bf v_+}$ is parametrized by the coefficients $f_n$.
Thus, the constraints on the choice of ${\bf v_+}$
are in one-to-one correspondence to
constraints on the choice of coefficients
$f_n$, namely
\begin{equation}
f_n\neq 0\mbox{~~forall}~~n,~~~
\sum_n\frac{f_n^* f_n}{1+t^2_n}=1~~~\mbox{and}~~
~\sum_n f_n^* f_n= \mbox{divergent .}
\label{fcond}
\end{equation}
Clearly, such $f_n$ only exist if $\vert
t_n\vert \rightarrow \infty$ for $n\rightarrow \infty$ and$/$or
for $n\rightarrow -\infty$. This reflects
the functional analytical fact that only
semibounded or fully unbounded
symmetric operators can have nonzero deficiency indices.

We can now use Eq.\ref{fbetr} to express the conditions on $\bf v_+ \rm , $ as
conditions on the coefficients $\{t^\prime_n\}$. We obtain as the constraints
on the data $\{t_n\}$ and $\{t^\prime_n\}$:
\be
t_{n+1}>t_n~, \mbox{~~and~~} t^\prime_n >0\mbox{~~for all~~} n ,
\ee
\be
\vert t_n\vert \rightarrow \infty \mbox{~~for~~}
n\rightarrow +\infty \mbox{~~and$/$or for~~}
 n\rightarrow -\infty
\ee
 and
\begin{equation}
\sum_n\frac{t^\prime_n}{1+t_n^2}=\mbox{~~finite},
~~~~~ \sum_n t^\prime_n=\mbox{~~divergent}.
\end{equation}

\subsection{Realness and differentiability of the
sampling kernel}
While we are working in a  complex Hilbert space, the aim in engineering
would be to sample real-valued signals. In the case of physical fields, we
may want to work with charged, i.e., complex-valued scalar fields, but we
will certainly also want to be able to work with real-valued scalar
fields.

We therefore have to deal with the fact that the representation of vectors
${\bf \phi}$ of the Hilbert space as functions $\phi(t)=( \ve,{\bf \phi})$
a priori depends on the arbitrary choice of the phases of the vectors
$\ve$. Correspondingly, also the sampling kernel $( \ve, \ve_n)$ depends
on the choice of these phases. For the sampling of real-valued signals we
therefore need to show that  it is possible to choose the phases of the
vectors $\ve$ such that if a signal is real-valued on one sampling lattice
$\{t_n\}$, then it is real-valued for all $t$.  This means that we need to
show that it is possible to choose the phases such that the sampling
kernel is real-valued.

In order to derive the appropriate choice of phases,
we reconsider the sampling kernel Eq.\ref{samker}:
\be
\left({\ve}_n, \ve\right)~ =~  \frac{
f_n ~ N(t)
}{t - t_n}
\label{samker2}
\end{equation}
which holds for all $t\neq t_n$, while, of course,
\be
\left(\ve_n,{\ve_n}\right)~ =~1~.
\label{normker}
\ee
We observe that the denominator in Eq.\ref{samker2}
goes linearly to zero whenever $t$ approaches one of the
$t_n$ of the lattice. Nevertheless, the absolute value of the
sampling kernel does not diverge since, as follows from Eq.\ref{Ne},
simultaneously also the numerator linearly approaches zero as $t$ approaches
$t_n$.

In particular, whenever $t$ passes through a $t_n$
the denominator in Eq.\ref{samker2} changes sign.
Thus, in order for the sampling kernel to be continuous,
it is necessary that also $N(t)$ changes sign at the points $t_n$, i.e.
as it passes through a zero.

Indeed, the definition of $N(t)$ in Eq.\ref{e19}
shows that it is possible to choose the
phases of the vectors $\ve \neq \ve_n$ such that
\be
N(t_r(\alpha)) ~=~(-1)^r \vert N(t_r(\alpha))\vert~.
\label{c1}
\ee
This choice of phases ensures that the sampling
kernel is real and continuous for all
$t\neq t_n$, and that, for all $n$,
both limits of the sampling kernel $(\ve_n,\ve)$ for
$t\rightarrow t_n$, from above and below, coincide.
However, in order to obtain that the sampling kernel
is continuous at the times $t_n$,
we still need to ensure that
those limits also coincide with the value $(\ve_n,\ve_n)=1$  set
by the normalization condition Eq.\ref{normker}.
Therefore, since $N(t)$ changes sign at each
$t_n$, also the $f_n$ must have alternating signs:
\be
f_n ~=~ (-1)^n ~\vert f_n\vert
\label{c2}
\ee
We still have the freedom to choose the phases of the vectors $\ve_n$.
Indeed, using  Eq.\ref{vp}, we can choose the phases of these
vectors  so that Eq.\ref{c2} holds.

We remark that the choice of phases to obtain the sampling kernel
is unique up to a global phase:
The sampling kernel is a scalar product of the vectors $\ve$ and therefore
the same real and differentiable sampling kernel is obtained if all vectors $\ve$
are multiplied by some constant overall phase $c$. Indeed, the definitions of
$N(t)$ and $f_n$ show that
this amounts to multiplying $N(t)$ with $c$ and multiplying $f_n$ with $c^*$,
so that the phase cancels out in the expression for the sampling kernel.
\sn
We have therefore shown that and how the phases of the vectors $\ve$ can be chosen
so that the sampling kernel is real and continuous, to obtain:
\be
\left({\ve_n}, {\ve_r(\alpha)}\right)~ =~  (-1)^{n+r}~
\frac{\vert f_n\vert ~ \vert N(t_r(\alpha))\vert }{t_r(\alpha) - t_n}
\end{equation}
The sign factor can be written, equivalently, as
\be
(-1)^{z(t_r(\alpha),t_n)}
\ee
where $z(t_r(\alpha),t_n)$ is the number of sampling points
$t_i$  in the interval $[t_r(\alpha), t_n]$ if $t_r(\alpha)<t_n$ or the number
of $t_i$ in the interval $[t_n,t]$ if
$t_n<t_r(\alpha)$, and $z=0$ if $t_n=t_r(\alpha)$.
\sn
Since each $t\in \R$ occurs in the family of lattices $\{t_r(\alpha)\}$
we can now express the sampling kernel in the general form:
\be
\left({\ve_n}, t\right)~ =~  (-1)^{z(t,t_n)}~
\frac{
\vert f_n\vert ~ \vert N(t)\vert }{t - t_n}
\label{samker3}
\end{equation}
Finally, using or results Eqs.\ref{fbetr},\ref{e19},
which express $\vert f_n\vert$ and $\vert N(t)\vert$
in terms of the data $\{t_n\}$ and $\{t_n^\prime\}$
we obtain:
\begin{equation}
\left({\ve_n}, \ve \right)~ =~ (-1)^{z(t,t_n)} \frac{\sqrt{t_n^\prime}}{t-t_n}
\left(\sum_m \frac{t_m^\prime}{(t-t_m)^2}\right)^{-1/2}
\end{equation}
\subsection{Calculating the full sampling kernel $G(t,t^\prime)$, which
yields all sampling lattices and
$\omega_{max}(t)$ for all $t$}
\label{fullker}
So-far we have only calculated the sampling kernel $(\ve,\ve_n)$
for one sampling lattice $\{t_n\}=\{t_n(0)\}$ of the family of sampling lattices.
It is straightforward to calculate the general sampling kernel:
\begin{eqnarray*}
(\ve_r(\alpha),\ve_s(\alpha^\prime))  & = & \sum_n ~
(\ve_r(\alpha), \ve_n)~(\ve_n, \ve_s(\alpha^\prime))\\ \\
 & = & \sum_n
\frac{(-1)^{r+s}~~~ t_n^\prime}{(t_r(\alpha)-t_n)(t_s(\alpha^\prime) -t_n)
\sqrt{
\left(\sum_m \frac{t_m^\prime}{(t_r(\alpha)-t_m)^2}\right)
\left(\sum_m \frac{t_m^\prime}{(t_s(\alpha^\prime)-t_m)^2}\right)}}
\end{eqnarray*}
or, in simplified notation, for arbitrary times $t, \tau \in \R$:
\be
(\ve,\bf \tau \rm )  ~=~ \sum_m
\frac{(-1)^{z(t,\tau)}~~~ t_n^\prime}{(t-t_n)(\bf \tau \rm -t_n)
\sqrt{
\left(\sum_m \frac{t_m^\prime}{(t-t_m)^2}\right)
\left(\sum_m \frac{t_m^\prime}{(\bf \tau \rm -t_m)^2}\right)}}
\label{fsk}
\ee
The significance of this equation is that, given one sampling lattice
$\{t_n\}=\{t_n(0)\}$ and its derivative
$\{t_n^\prime\}=\{\frac{d}{d\alpha} t_n(0)\}$, Eq.\ref{fsk} allows us
reconstruct the entire family $\{t_n(\alpha)\}$ of sampling lattices, up
to reparametrization of $\alpha$. This is because each of the sampling
lattices consists of the eigenvalues of orthogonal vectors, i.e. of
vectors whose scalar product vanishes:

For any $t_0\in \R$ there exists  a self-adjoint
extension $T(\alpha)$ and some index $n$ so that
$t_0=t_n(\alpha)$. The entire sampling lattice $\{t_n(\alpha)\}$ then consists of
$t_0$ and all $t$ for which $(\ve_0, \ve)=0$:
\be
\{t_n(\alpha)\}~=~ \{t_0\} ~\cup~ \{ t~~\vert~~ (\ve_0, \ve)=0\}
\ee
To this end, for generic data $\{t_n\}$ and $\{t_n^\prime\}$ the zeros of
the general sampling kernel in Eq.\ref{fsk} can at least be
found numerically. Thus, all lattices of the family $\{t_n(\alpha)\}$ can
 be calculated, in principle, from
the data $\{t_n\},\{t_n^\prime\}$ which enter the sampling theorem.

In turn, once all sampling lattices are known, also the curve
$\omega_{max}(t)$ can be constructed explicitly for all $t$, as explained
above.

\subsection{Other sampling lattices}
In the new sampling theorem, the class of signals to be sampled is chosen
by specifying  one sampling lattice  $\{t_n\}$ and its derivative,
$\{t_n^\prime\}$ and by thereby specifying the time-varying bandwidth
curve $\omega_{max}(t)$. The sampling theorem allows us to recover the
signals of this class from their values at any one of the sampling
lattices from the family of sampling lattices which is determined by the
data $\{t_n\}$ and $\{t_n^\prime\}$.

In fact, the same class of signals, with bandwidth specified by the data
$\{t_n\}$ and $\{t_n^\prime\}$, can also be sampled on a large variety of
other sampling lattices, by using a suitably modified sampling formula.

To this end, assume that $\{\tilde{t}_n\}$ is any sampling lattice for
which there exists a linear transformation $K$, such that:
\be
\sum_r~K_{sr}~(\tilde{\ve}_r,\ve_m) = \delta_{sm}
\ee
Applied to
\be
\sum_m ~(\tilde{\ve}_r,\ve_m)~(\ve_m,\phi) ~=~ (\tilde{\ve}_r,\phi)
\ee
this yields
\begin{eqnarray*}
(\ve_n,\phi)& = & \sum_{m,r}~K_{n,r}~(\tilde{\ve}_r,\ve_m)~(\ve_m,\phi) \\
 & = & \sum_r~ K_{n,r}~(\tilde{\ve}_r,\phi)
\end{eqnarray*}
Inserting this result into $(\ve,\phi) = \sum_n~(\ve,\ve_n)~(\ve_n,\phi)$, we
obtain the sampling formula for sampling on the lattice $\{\tilde{\ve}_n\}$:
\be
(\ve,\phi) = \sum_{n,r}~(\ve,\ve_n)~K_{n,r}~
(\tilde{\ve}_r,\phi)
\ee
\section{Summary of the new sampling method}
\label{secsummary} The new sampling method allows us to continuously
increase and lower the sampling rate over time in order to sample signals
with a correspondingly time-varying bandwidth. For engineering purposes,
this allows one to economize on data collection whenever the bandwidth is
temporarily low and to increase the sampling rate whenever the bandwidth
is temporarily high. This is the basis of the data compression method
patented in \cite{ak-patent}.

The mathematical background is based on the observation that the action of
the ``time" operator $T: T\phi(t)=t\phi(t)$ can only infinitesimally
change the spectrum of bandlimited signals (because it acts as
differentiation in Fourier space), independently of the chosen bandwidth.
We were able, therefore, to identify classes of signals with time-varying
bandwidth as analytic domains of simple symmetric multiplication operators
$T: \phi(t) \rightarrow t\phi(t)$. (In the case of fields in spacetime, we
may think, equivalently, of $T$ as a position coordinate.)

In particular, we found that, as in the case of constant bandwidth, also
in the case of a time-varying bandwidth, there exists a family of lattices
$\{t_n(\alpha)\}$ of sampling times at which the signals can be measured
and recovered for all $t$. We found that the lattice spacings of these
sampling lattices, together, form a well-defined function of time $s(t)$.
This yielded a precise definition of time-varying bandwidths as
time-varying Nyquist rates, i.e. as a time-dependence of the spacings
$s(t)$ of sampling lattices. We defined $\omega_{max}(t)=1/2s(t)$.
\sn
We also discussed that, contrary to intuition, a finite bandwidth does not
imply that signals vary ``at most as fast as their highest Fourier
component". We found that instead a true property of finite bandwidth
signals is that they obey a lower bound to how much they can be peaked, in
the sense that the standard deviation of the samples $\phi(t_n(\alpha))$
(for arbitrary fixed $\alpha$) of strictly bandlimited signals is bounded
 from below.

More generally, we found that classes of signals with time-varying
bandwidths obey a time-varying lower bound on their standard deviation,
i.e. they consist of signals which, if centered around time $t$, have
standard deviations bounded from below by a positive time-dependent
function $\sigma_{max}(t)$.
\subsection{How to use the sampling theorem}
Concretely,  the sampling theorem  requires as input data first the choice
of one sampling lattice $\{t_n\}$. The choice of one sampling lattice
largely determines the Nyquist rate and therefore the time-varying
bandwidth $\omega_{max}(t)$. This is because by giving one sampling
lattice the Nyquist rate $s(t)$, i.e. the lattice spacing, is defined
already for a discrete set of intervals. This means that the bandwidth is
already defined at a discrete set of times, namely at the times $\tau_n
=(t_n+t_{n+1})/2$, i.e. at the center of each lattice interval, through
the inverse of the spacing:
\begin{equation}
\omega_{max}(\tau_n) ~=~ \frac{1}{2s(t)}~=~ \frac{1}{2~(t_{n+1}-t_n)}
\label{o1}
\end{equation}
We also found that the bandwidth cannot vary much between these points:
\be \left\vert \frac{d}{dt} ~
\omega_{max}(t)\right\vert < 4~ \omega_{max}(t)^2. \ee
The sampling theorem requires a further set of input data,
$\{t_n^\prime\}$, to fine-tune the bandwidth curve by specifying the
derivative of the bandwidth at those points, namely through:
\begin{equation}
\frac{d}{dt}~\omega_{max}(\tau_n)~=~
-4~\omega_{max}(\tau_n)^2~\frac{
 (t^\prime_{n+1}- t^\prime_n)}{ (t^\prime_{n+1}+t^\prime_n)}
\label{od}
\end{equation}
We showed that the bandwidth curve $\omega_{max}(t)$ thereby becomes
determined for all $t$. We recall that by embedding the sampling lattice
$\{t_n\}$ as the lattice $\{t_n(0)\}$ into the family of sampling lattices
$\{t_n(\alpha)\}$ which is determined by (and determines) the bandwidth
$\omega_{max}(t)$, the auxiliary data can be identified as the derivative
of the lattice: $\{t_n^\prime\}= \{d/dt ~t_n(\alpha)\vert_{\alpha=0}\}$.
\sn
After choosing sampling times $\{t_n\}$, which already largely determine
the bandwidth, and auxiliary data $\{t_n^\prime\}$, which fine-tune and
fix the bandwidth curve $\omega_{max}(t)$ for all $t$, the sampling
theorem then allows to recover all signals of the class with this
bandwidth through the sampling formula:
\begin{equation}
\phi(t) ~ = ~ \sum_n ~G(t,t_n) ~~\phi(t_n)
\label{ses1}
\end{equation}
with the sampling kernel $G(t,t_n)$:
\begin{equation}
G(t,t_n) = (-1)^{z(t,t_n)} \frac{\sqrt{t_n^\prime}}{t-t_n}
\left(\sum_m \frac{t_m^\prime}{(t-t_m)^2}\right)^{-1/2}
\label{mr}
\end{equation}
The number of eigenvalues $t_n$ in a given interval defines what we call
the local density of degrees of freedom and variable bandwidth. Taking
samples at the so-defined Nyquist rate expands functions in terms of an
orthonormal basis of functions. The reconstruction is, therefore, stable
in the sense that small perturbations in the sample values lead to small
perturbations in the reconstructed function. Stability of the
reconstruction is important since sample values are generally subject to
noise and/or to quantum uncertainties. For the case of conventional
Shannon sampling it is known that reconstruction from samples taken at a
lower rate than the Nyquist rate is possible, in principle, but must be
unstable, as Landau proved in \cite{landau}.

Here too, we may choose to use the data $\{t_n\},\{t_n^\prime\}$ merely in
order to fix the bandwidth curve $\omega_{max}(t)$ and then decide to
sample the so-specified class of signals on a different sampling lattice.
Indeed, we found that the same class of signals can be sampled also on any
other lattice, say $\{\tilde{t}_n\}$, for which $K_{r,s}=
(G(\tilde{t_r},t_s))_{r,s}$ is invertible. The sampling formula then
reads:
\begin{equation}
\phi(t) = \sum_{n,m} G(t,t_n) ~K^{-1}_{n,m}~\phi(\tilde{t}_m)
\end{equation}
The spacing of these sampling lattices $\{\tilde{t}_n\}$ may be larger
than the spacing of the eigenvalues of the self-adjoint extensions of $T$.
Since we are covering conventional Shannon sampling as a special case, see
Sec.\ref{recovershannon}, it is to be expected that reconstruction is
necessarily unstable if the samples are taken at a lower rate than the
density of the degrees of freedom. It should be interesting to prove this
in generality.
\bigskip\newline
Finally, let us briefly address the case of engineering applications,
where all sampling lattices are finite.

We recall from Sec.\ref{nofin} that finite sampling lattices do not occur
in the theoretical framework. Technically, the reason is that the
underlying time operator $T$ can only be simple symmetric if it is
unbounded. Thus, in principle, in Eqs.\ref{ses1},\ref{mr} the indices $n$
and $m$ which label the sampling times $t_n$ must run through a countably
infinite index set, e.g. the natural numbers or the integers. Therefore, a
truncation error appears to be unavoidable.

Our generalized sampling theorem allows us, however, to freely choose
the sampling points $\{t_n\}$, as long as the sequence $\{t_n\}$ is divergent.
This freedom can be used
to sample on effectively finite sampling lattices, say $\{t_n\}_{n=1}^{n=N}$:

To this end, we  extend the finite sampling lattice $\{t_n\}_{n=1}^{n=N}$
on which we wish to sample to an infinite sampling lattice by adding an
infinite number of  further sampling points. These additional sampling
points we are free to choose for example at very very large $t$, where the
signals have negligible support. Thus, these sampling points do not
contribute to either of the sums in Eqs.\ref{ses1},\ref{mr}. In practise,
we can therefore leave Eqs.\ref{ses1},\ref{mr} unchanged and we simply
restrict the sums over $n$ and $m$ to the finite sampling lattice
$\{t_n\}_{n=1}^{n=N}$.

As discussed in Sec.\ref{effi}, the class of signals which is being
sampled in this case has a bandwidth curve $\omega_{max}(t)$ which decays
as $1/t$ outside the interval $[t_1,t_N]$ in which the samples are taken.

\subsection{Recovering conventional Shannon sampling as a special case}
\label{recovershannon} The standard Shannon Sampling Formula is to arise
as a special case. The class of signals with constant bandwidth
$\omega_{max}$ is characterized by the family of lattices
$\{t_n(\alpha)\}$ with $t_n(\alpha)= \alpha + n/2\omega_{max}$. Choose the
sampling lattice $\{t_n\}=\{t_n(0)\}$. Its derivative $\{t^\prime_n\}$ is
$t^\prime_n=1$. Thus, Eq.\ref{mr} yields the sampling formula:
$$
\phi(t) =
\left( \sum_{m=-\infty}^{+\infty} \frac{1}{(t-t_m)^2}\right)^{-1/2}
\sum_{n=-\infty}^{+\infty} \frac{(-1)^{z(t,t_n)}}{t-t_n}~~\phi(t_n)
$$
Since this equation must coincide with the Shannon formula, Eq.\ref{e1},
we can conclude that
$$
\frac{\pi^2}{\sin^2(\pi x)} =
\sum_{n=-\infty}^{+\infty}\frac{1}{(x-n)^2}
$$
which provides an example of a series
expansion deduced with the new method. Note
that setting $x=1/2$
we obtain the expansion $\pi =
\sqrt{\sum_{n=-\infty}^{+\infty}(n-1/2)^{-2}}$.

\section{On the relation to coherent states
 and wavelets}

\subsection{One - and two parameter overcomplete, continuous and normalizable
resolutions of the identity}
We are here obtaining
 resolutions of the identity
in terms of an overcomplete and  continuously
para\-metr\-ized set of normalizable vectors,
\begin{eqnarray}
1 & = & \frac{1}{2\pi}\int_0^{2\pi}d\alpha\sum_n
\ve_n^*(\alpha)\otimes
\ve_n(\alpha)  \nonumber \\
 & = & \frac{1}{2\pi}~\int_{-\infty}^{+\infty} dt~\frac{
d\alpha}{dt}~ \ve^*\otimes \ve \label{resid}.
\end{eqnarray}
In this sense, our expansion is similar to coherent states and wavelet
expansions. The main difference is that while we arrive at a one-parameter
resolution of the identity in the Hilbert space, coherent states and
wavelets, see e.g. \cite{jrk}, are typically characterized, analogously,
by two-parameter resolutions of the identity. In this context, it will be
interesting to follow the analogous construction to the one here, but for
time-operators with higher deficiency indices.
\sn
The resolution of the identity obtained in Eq.\ref{resid} implies that we
are in fact dealing with a reproducing kernel Hilbert space. Namely, the
signals ${\bf \phi}$ obey
\be
(\ve,\phi)  ~=~
\frac{1}{2\pi}~\int_{-\infty}^{+\infty}d\tau~\frac{d\alpha}{dt}(\tau)~
(\ve, \bf \tau \rm )~(\bf \tau \rm , \phi)
\ee
so that, with the reproducing kernel
\be
R(t,\tau)~:=~  (\ve, \bf \tau \rm )~ \frac{d\alpha}{dt}(\tau)
\ee
we obtain:
\be
\phi(t) = \frac{1}{2\pi}\int_{-\infty}^{+\infty} d\tau~R(t,\tau)~\phi(\tau)
\ee
\subsection{Calculation of the reproducing kernel}
In Sec.\ref{fullker} we calculated the sampling kernel $(\ve, \bf \tau \rm
)$ for all $t,\tau$ in terms of the input data $\{t_n\},\{t_n^\prime\}$.
In order to determine the reproducing kernel it remains to calculate
$d\alpha/dt$.  To this end, we solve Eq.\ref{co} for $\alpha$, by using
\be
\cot^{-1}(z) ~=~ \frac{i}{2}~ \ln~ \frac{iz+1}{iz-1},
\ee
so that
\be
\alpha ~=~i~\ln ~\frac{iy(t)+1}{iy(t)-1}
\ee
where
\be
y(t)~=~ \sum_n~t_n^\prime~\frac{t~t_n+1}{(t-t_n)(t_n^2+1)}~~.
\ee
We can now calculate
\be
\frac{d\alpha}{dt}~=~\frac{d\alpha}{dy}\frac{dy}{dt}
\ee
  from
\be
\frac{d\alpha}{dy}~=~ \frac{-2}{1+y^2}
\ee
and
\be
\frac{dy}{dt} ~=~ \sum_n t_n^\prime~\frac{1}{(t-t_n)^2}~~.
\ee
Thus, we obtain
\be
\frac{d\alpha}{dt} ~=~ \frac{2\sum_m t_m^\prime\frac{1}{(t-t_m)^2}}{1+\left(
\sum_n t_n^\prime\frac{t~t_n +1}{(t-t_n)(t_n^2+1)}\right)^2}~,
\ee
which finally yields a fully explicit expression for the reproducing
kernel:
\be
R(t,\tau) ~=~ (-1)^{z(t,\tau)}~~
\sqrt{\frac{\sum_r\frac{t_r^\prime}{(\tau-t_r)^2}}{
\sum_s\frac{t_s^\prime}{(t-t_s)^2}}}~~
\frac{\sum_n\frac{2~t_n^\prime}{(t-t_n)(\tau- t_n)}}{
1+ \left(\sum_m\frac{t_m^\prime~(\tau~t_m +1)}{(\tau-t_m)(t_m^2+1)}\right)^2   }
\ee
\section{Outlook}

In the literature, studies of quantum gravity and string theory, see
\cite{qg}, have suggested a natural ultraviolet cutoff in the form of a
finite lower bound to the resolution of distances with a corresponding
modification of the uncertainty relations. As discussed in Sec.\ref{five},
theories that are ultraviolet cutoff in the sampling theoretic sense
concretely realize such generalized uncertainty relations. A sampling
theoretic ultraviolet cutoff at the Planck scale also corresponds to a
finite density of degrees of freedom for physical fields in the very same
sense that bandlimited music signals possess a finite density of degrees
of freedom and a correspondingly finite information density.

This leads to the prospect that sampling theory could allow us to
formulate physical theories which combine the virtues of lattice theories
with the virtues of continuum theories. Fields, such as those of quantum
field theoretic path integrals, could be defined over a continuous
spacetime manifold or, entirely equivalently, over any lattice which is
sufficiently densely spaced. A theory's ultraviolet finiteness would be
displayed when formulated on one of the lattices, while the same theory's
preservation of external symmetries would be displayed when written as a
continuum theory.

Here, we proved sampling results for the one-dimensional case as a
preparation for the generalization of Shannon sampling theory to generic
curved spacetimes which has been outlined in \cite{ak-s} and
\cite{ak-covsam}. To this end, we generalized sampling theory using the
functional analysis of simple symmetric operators which have deficiency
indices $(1,1)$ and whose self-adjoint extensions have purely discrete
spectra. It should be straightforward to apply these functional analytic
methods to simple symmetric operators with higher deficiency indices,
thereby leading to sampling methods for vector- and spinor-valued fields.
The very same methods should be useful for developing the details of the
general sampling theory of fields on n-dimensional curved spacetimes.

It is encouraging that the mathematical language of the sampling
theoretical Planck scale cutoff, namely functional analysis, is also the
language of quantum theory.
\mn \bf Acknowledgement: \rm The author is happy
to thank John Klauder for useful criticisms.

\end{document}